# Phase determination in spin-polarized neutron specular reflectometry by using a magnetic substrate

S. Farhad Masoudi*,[1]

[1]Physics Department, University of Tehran, P.O. Box 1943-19395, Tehran, Iran

The scattering length density profile for a thin film structure can be determined uniquely if both modulus and phase of the reflection coefficient is known. Here, we describe a method for recovering the phase information which utilize a magnetic substrate and based on polarization analysis of the reflected beam. The method is derived in the formalism of transfer matrix so it is applicable for any unknown real scattering length density of nonmagnetic films (i.e. in the case where there is no effective absorption).

## I. Introduction

Neutron specular reflectometry has become a suitable technique for investigation of the physics of many surface and interfacial structures [1,2]. The measurement of the specularly reflected intensity in term of the wave number q perpendicular to the sample surface provides important information on the atomic or magnetic scattering length density (SLD) profile of the nanostructure materials along their depth x [3]. However, extracting the profile from the measured reflectivity R(q), as a function of q, has been hampered by the so called phase problem [4,5]. This problem refers to the loss of the phase of reflection coefficient in any scattering problem. In the absence of the phase, least-squares-fit methods allow the determination of the depth profile [6,7] but in general the solution is not unique since different profiles may produce the same reflectivity [8,9]. If the reflection coefficient is known in modulus and phase, it can inverted for the depth profile in a fairly straightforward way, by solving the Gelfand-Levitan-Marchenko integral equation or practical algorithms have been developed [10-12]. Several solutions of phase problem have been proposed for neutron specular reflection, including these devised and cited in Ref. [13-20]. Nonetheless, only the reference layer [13, 14] and variation of surrounding method [15] have proven to be experimental practical. The reference layer methods based on polarization measurements [18, 19] are of particular interest, because they also work in total reflection regime and allow reconstruction of surface profiles of magnetic or absorptive non magnetic samples [20]. Recently this method has been derived in the formalism of the transfer matrix [21].

In this brief report we show a method of phase determination by using a magnetic transmitting media (substrate) and fully polarization analysis of the reflected beam. One of the advantages of the method is that it is not necessary to change the substrate and only two scalar value of the nuclear and magnetic SLD for magnetic substrate media are needed.

## II. Method

To derive the phase determination method, we need the relation between the polarization of the reflected beam for an arbitrary finite film and the magnetic substrate (It is assumed that the incident medium is vacuum). We assume that the substrate is magnetized in a direction parallel to the reflecting surface, z, which is taken as the direction of spin quantization. For a magnetic film, when the magnetization is in the plane of the film, the SLD depth profile is proportional to $\rho(x)\pm\mu B$, where $\rho$ is the SLD at a depth x from the surface determined by the product of the average number scattering amplitude and the atomic number density, B is the magnetic field, and $\mu$ is a material constant. The two signs for the second term refer to neutron polarized parallel and opposite to the local magnetization, respectively. The magnetization of the magnetic substrate will also generate a magnetic induction outside the ferromagnetic film, which we assume, however, to be small enough not to affect the neutron beam. In this case if the incident beam polarized in x direction, the polarization of the reflected beam can be expressed in terms of the reflection coefficient $r_{\pm}(q)$ (subscripts '+' and '-' denote to

*Corresponding author.
E-mail: fmasoodi@chamran.ut.ac.ir.

incident neutron beam polarized parallel and antiparallel to the magnetic field respectively) as follow:

$$P_x + iP_y = \frac{2r_+^* r_-}{R_+ + R_-} \tag{1}$$

$$P_z = \frac{R_+ - R_-}{R_+ + R_-}, \tag{2}$$

where $R_\pm(q)=|r_\pm|^2$.

If the SLD of substrate is referred to $\rho_+$ and $\rho_-$ for plus and minus magnetization of the substrate, respectively, $r_\pm(q)$ can be written as

$$r_\pm(q) = \frac{\beta_\pm - \alpha_\pm - 2i\gamma_\pm}{\alpha_\pm + \beta_\pm + 2}, \tag{3}$$

where

$$\alpha_\pm = h_\pm A^2 + h_\pm^{-1} C^2$$
$$\beta_\pm = h_\pm B^2 + h_\pm^{-1} D^2 \tag{4}$$
$$\gamma_\pm = h_\pm AB + h_\pm^{-1} CD$$

and where

$$h_\pm = (1 - 4\pi\rho_\pm / q^2)^{1/2}, \tag{5}$$

The four real functions A, B, C and D are uniquely determined by the SLD of the film and are the elements of the 2×2 transfer matrix which carries the exact wave function and its first derivative across the film, from front edge to back. Explicitly,

$$\begin{pmatrix} 1 \\ ih_\pm \end{pmatrix} t_\pm e^{ih_\pm qL} = \begin{pmatrix} A & B \\ C & D \end{pmatrix} \begin{pmatrix} 1 + r_\pm \\ i(1 - r_\pm) \end{pmatrix}, \tag{6}$$

where the column matrices contain the transmission and reflection coefficients, $t_\pm$ and $r_\pm$, characterizing the wave function and its derivative in the transmitting and incident media, respectively, and L is the film thickness. The directly measured reflectivity, R(q), depends on these characteristic functions in terms of a new defined quantity, $\Sigma(q)$, as

$$\Sigma_\pm(q) = 2\frac{1 + R_\pm}{1 - R_\pm} = \alpha_\pm + \beta_\pm \cdot \tag{7}$$

we can see at once from Eqs. (3) and (7) that reflectivity contains less information than reflection coefficient, an alternative perspective of the phase problem.

By using Eqs. (1), (2) and (3) to find the polarization in three directions as a function of the transfer matrix elements, after some straightforward algebra, we obtain

$$P_x = 1 - 2\frac{(h_+ - h_-)^2}{h_+ h_-}\frac{1}{\Sigma_+\Sigma_- - 4}, \tag{8}$$

$$P_y = 2\frac{(h_+^2 - h_-^2)}{h_+ h_-}\frac{\tilde{\gamma}}{\Sigma_+\Sigma_- - 4}, \tag{9}$$

$$P_z = 2\frac{\Sigma_+ - \Sigma_-}{\Sigma_+\Sigma_- - 4}, \tag{10}$$

where the tilde denotes a reversed film, which corresponds to interchange of the diagonal elements of transfer matrix (i.e. A,B,C,D→d,b,c,a in Eq. (6)).

Eqs. (8)-(10) are the essential formulas for our method. Eqs. (8) and (9) show that $\tilde{\gamma}$ can be determined by $P_x$ and $P_y$ as follow

$$\tilde{\gamma} = \frac{h_+ - h_-}{h_+ + h_-}\frac{P_y}{1 - P_x}. \tag{11}$$

Two others unknown parameters can be determined by $\Sigma_+$ and $\Sigma_-$. $\Sigma_\pm$ can be obtained using Eqs. (8) and (10) by following quadratic equation

$$\Sigma_\pm^{\;2} \pm m\Sigma_\pm = (4 + m\frac{1 - P_x}{P_z}), \tag{12}$$

where

$$m = \frac{P_z}{P_x - 1}\frac{(h_+ - h_-)^2}{h_+ h_-}. \tag{13}$$

This equation has two different solutions and the physical solution must be selected which satisfies the condition $\Sigma_\pm \geq 2$. The numerical examples show that only one of these solutions satisfies the physical condition, however, we don't have any general proof for it. Once the physical solution of equation 12 is found, two other unknown parameters are obtained as follow

$$\tilde{\alpha} = h_+ h_- \frac{h_+\Sigma_- - h_-\Sigma_+}{h_+^2 - h_-^2}, \tag{14}$$

$$\tilde{\beta} = \frac{h_+\Sigma_+ - h_-\Sigma_-}{h_+^2 - h_-^2}, \tag{15}$$

After determination of $\tilde{\alpha}$, $\tilde{\beta}$ and $\tilde{\gamma}$, the complex reflection coefficient of the reversed unknown film is known. Note from Eq. (5) that $\tilde{r}(q)$ can not be determined directly at q below the critical $q_c$, where $q_c^2=4\pi\rho_+$. The missing values below $q_c$ can be accurately reconstructed by using the fact that the reflection coefficent approaches to (-1) when incident wave number approaches to zero. However, since in our method the absorption is neglected, $\tilde{r}(-q) = \tilde{r}^*(q)$ and so $\tilde{r}(q)$ is known over the entire q-domain. Once $\tilde{r}(q)$ has been obtained, by solving the Gel'fand-Levitan-Marchenko integral equation or related relations [10-12], it can be inverted for the SLD of reversed unknown sample. Thus, once we have found this SLD, we simply need to find the mirror image of the result to retrieve the free film SLD profile.

## III. Example

As a realistic example to test the method by simulation, we consider a sample with a 30nm thick gold on a 20nm Cr having constant SLD value of $4.46\times10^{-4}$ and $3.03\times10^{-4}$ $nm^{-2}$ respectively which is mounted on top of a thick Co layer having SLD values of $2.23\times10^{-4}nm^{-2}$ respect to non magnetization as the magnetic substrate. The SLD of the substrate increases or decreases respect to plus or minus magnetization. The magnetization of the substrate will also generate a magnetic induction outside the ferromagnetic film, which we assume to be small enough not to affect the neutron beam. For simplicity, we set it equal to zero. The substrate is magnetized in a direction parallel to the reflecting surface (+z direction) which is taken as the direction of spin quantization. It is assumed that the magnetic field vanishes in the unknown sample. Absorption and roughness of interfaces are neglected.

The q dependence of the polarization of reflected beam, $P_x$ ,$P_y$ and $P_z$ for incident beam to be fully polarized in the +x-direction has been shown in Fig. (1). In all calculation the simulated data start at the critical q of the substrate.

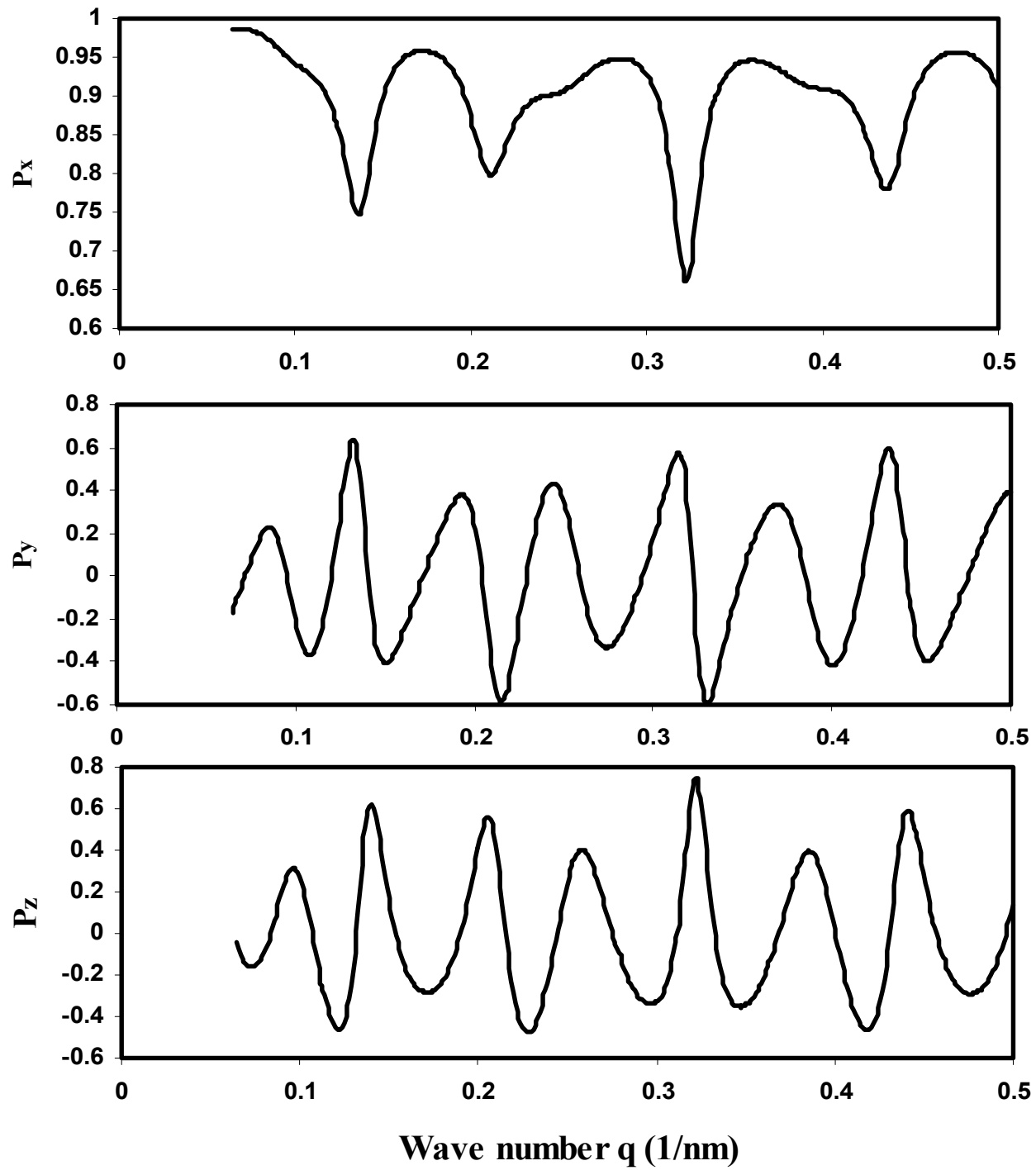

Fig. 1. Simulated polarization data for our example. The incident beam is assumed to be fully polarized in x direction. The simulated data start at the critical q of the substrate.

By using the measurements of $P_x$ and $P_z$ components in Eq. (12), we obtain two solutions for each $R_+$ and $R_-$, Fig. (2). As it is seen, only one of two solutions satisfy $R_\pm(q)\leq 1$ and is therefore physically admissible. Using the reconstructed data of the reflectivity and the measurements of $P_y$ components the real and imaginary part of the complex reflection coefficient of the reversed unknown film (i.e. the Cr layer mounted on top of the Au layer) can be extracted by following the procedure outlined in Sec. II which are displayed in Figs. (3).

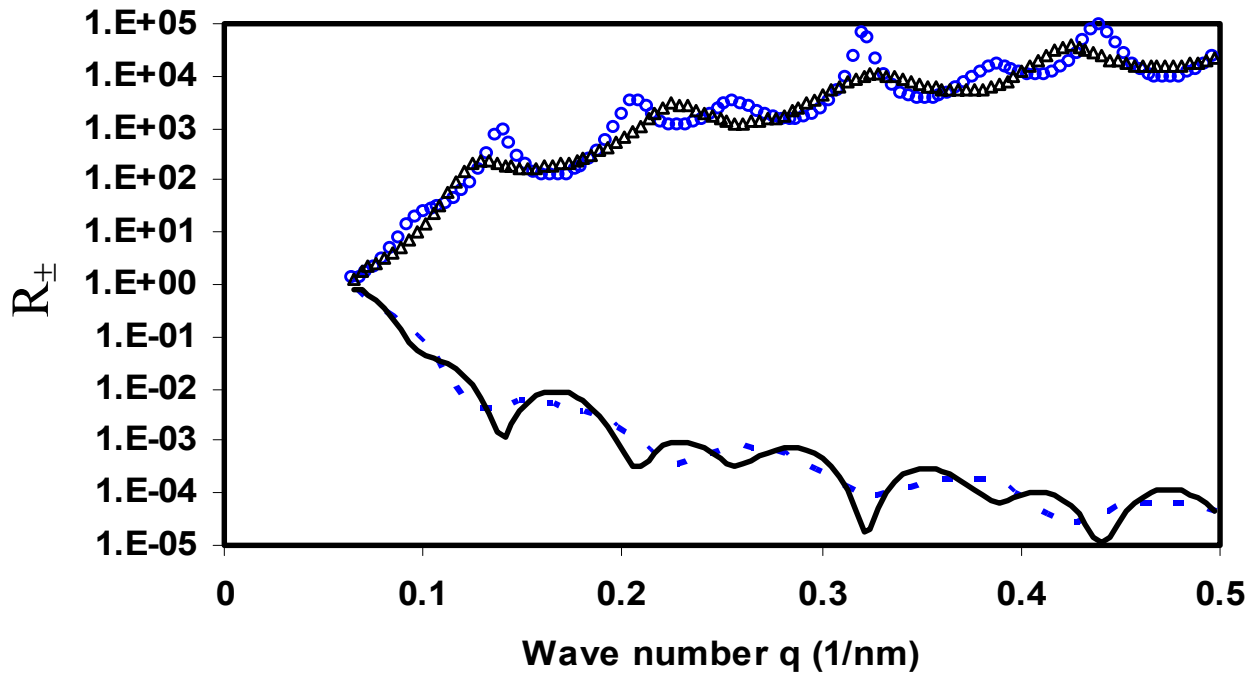

Fig. 2. $R_+$ and $R_-$ respect to two solutions of the quadratic equation as obtained from the measurements of $P_x$ and $P_z$. As it is seen, only one of the solutions satisfies $R_\pm\leq 1$ respect to $\Sigma_\pm\geq 2$. (Dash line and circles for $R_+$ and solid line and triplets for $R_-$).

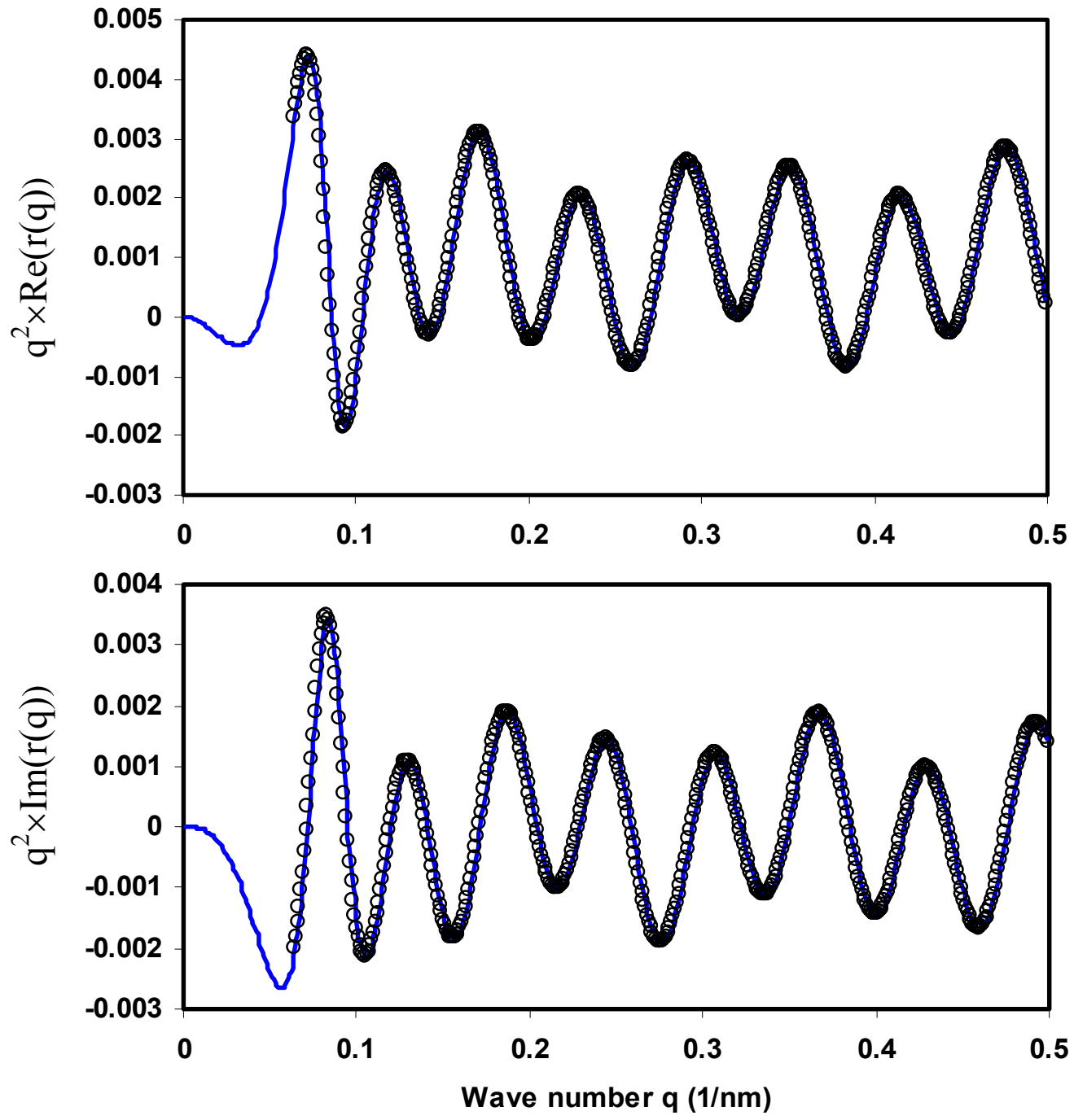

Fig. 3. $q^2\times \tilde{r}(q)$ for 30 nm thick Au layer mounted on 20 nm Cr layer (a: Real part and b: Iamginary part). Solid line: Computed directly from Eq. (3), Circles: Recovered from the measurements of the polarization components for q greater than the critical q of the substrate.

## Conclusions

We have proposed a method involving a magnetic substrate which allows a determination of the full complex reflection coefficient for any unknown nonmagnetic films in the case where there is no effective absorption. Thus a reconstruction of surface profiles of any unknown real potential becomes feasible. The method is based on polarization analysis of the reflected beam instead of the reflectivity. Using a magnetic substrate has a significant advantage that only two single scalar value of the SLD for up and down magnetization of the substrate is needed, in contrast to having to accurately know the SLD profiles of buried reference layers. On the other hand, against the surround method, in this method it is not necessary to change the surrounding media, however polarized neutron beams instruments are required. As in this method both the modulus and the phase of the reflection coefficient can be extracted, it is applicable for supported potentials with bound states, i.e., potentials of finite thickness that are negative.